\documentclass[11pt]{article}
\usepackage{fleqn,cospar}

\usepackage{url}

\usepackage{graphicx}
\usepackage[figuresright]{rotating}

\title{RADIO GALAXIES AND THE ACCELERATION OF THE UNIVERSE 
BEYOND A REDSHIFT OF ONE}

\author{Ruth A. Daly\address{Berks-Lehigh Valley College, 
Penn State University, Reading, PA 19610, USA}}

\begin{document}

\maketitle

\begin{abstract}
Radio galaxies provide a means to determine the coordinate distance,
the luminosity distance, the dimensionless luminosity distance, or
the angular size distance  
to sources with redshifts as large as two.  Dimensionless 
coodinate distances for 55 supernovae and 20 radio galaxies 
are presented and discussed here.  The radio galaxy results are
consistent with those obtained using supernovae,  
suggesting that neither method is plagued by unknown 
systematic errors. The acceleration parameter q(z) 
and the expansion rate H(z) or dimensionless expansion
rate E(z)  
can be determined directly from the data without having
to make assumptions regarding the nature or evolution of the
``dark energy.''  The expansion rate E(z) can be determined from 
the first derivative of the dimensionless coordinate distance, 
$(dy/dz)^{-1}$, and the acceleration parameter
can be determined from a combination of the first 
and second derivatives of the
dimensionless coordinate distance. A model-independent 
determination of E(z) will allow the properties and 
redshift evolution of the ``dark energy'' to be
determined, and a model-independent determination of 
q(z) will allow the redshift at which
the universe transitions from acceleration to deceleration
to be determined directly. Determination of E(z) and
q(z) may also elucidate 
possible systematic errors in the determinations of
the dimensionless coordinate distances.  
 
\end{abstract}

\section*{INTRODUCTION}

A primary goal of current cosmological studies 
is to determine whether the expansion of the universe 
is accelerating or 
decelerating at the present epoch, and what the acceleration or 
deceleration rate is.  From this, we can determine the 
global cosmological parameters, and study how structure evolved.  

One way to improve our understanding of the recent history of 
the universe, and begin to quantify 
the properties of the ``Dark Energy," 
is to observationally determine the redshift at 
which the universe transitions from acceleration to deceleration 
if is is established that the expansion of the universe
is accelerating at the present epoch.   ``Dark Energy''
is discussed, for example, by Caldwell, Dave, \&
Steinhardt (1998), Turner \& Riess (2002), Frieman,
Huterer, Linder, \& Turner (2002), 
Peebles \& Ratra (2002), 
and Peebles (2002).

Since astrophysics is a ``spectator science," 
for which we can only collect data but
cannot design and control experiments, 
unknown systematic errors are always a worry.  
The only way to convincingly establish a result 
is to have independent determinations that agree.

\section*{THE BASICS}

The Robertson-Walker metric describes an 
expanding (or contracting) homogeneous and 
isotropic space-time, and has the line element

\begin{equation}
{d \tau^2 = dt^2 - a^2(t) 
\left( {dr^2 \over 1-kr^2} +r^2 d\theta^2 + r^2~sin^2\theta ~d\phi^2 \right)}
\end{equation}
(see, for example, Weinberg 1972), where the cosmic 
scale factor $a(t)$ is related to the source redshift $z$
and the current value of the cosmic scale factor $a_o$:
$(a(t)/a_o) = (1+z)^{-1}$.  For light traveling from 
a source with redshift $z$ along a radial path, equation
(1) implies that the coordinate distance to the source,
$a_or$ can be obtained by integrating the equation
\begin{equation}
dr/\sqrt{1-kr^2} = dt/a(t)~.
\end{equation}
In this universe, a source at redshift z, 
with intrinsic physical size D, 
and luminosity L, will have an observed angular size $\theta$
given by
$\theta = D (1+z) / (a_or)$ $= D / d_A$, where  
$d_A$ is the ``angular size distance" 
$d_A = (a_or)/(1+z)$ and 
$(a_or)$ is the coordinate distance to the source.  
The flux $f$ that could be detected from the source is given by
$f = L/[4 \pi (a_o r)^2 (1+z)^2]$ $= L/[ 4\pi d_L^2]$, where 
$d_L$ is the ``luminosity distance" 
$d_L = (a_or)(1+z)$ $= (H_o)^{-1}  y(z) (1+z)$; 
$y(z) = H_o a_o r$ is the dimensionless coordinate distance
(see, for example, Peebles 1993).

Thus, if $d_A$ or $d_L$ to a source at redshift z can be 
observationally determined, then the coordinate distance 
$(a_or)$ and the dimensionless coordinate distance 
$y(z)$ to that redshift are known.    

The coordinate distance is related to the cosmological parameters 
through the equations $a_o \int dr/\sqrt{1-kr^2} = \int (a/ \dot{a}) dz$,
and 
$(\dot{a}/a) = H_o E(z)$.  For $k=0$, the coordinate
distance is $(a_or) = H_o^{-1} \int dz/E(z)$.  
For a universe with components for which the 
equations of state $w_i$ are time independent,
such as a universe with quintessence (Caldwell, 
Dave, \& Steinhardt 1998)
$E^2(z) = \sum \Omega_i(1+z)^{n_i}$, 
where
$w_i=P_i/\rho_i$, and
$n_i =3(1+w_i)$ for a component with non-evolving equation of
state (see, for example, the Appendix of Daly \& Guerra 2002).  The  
deceleration parameter at the present epoch is 
$q_o= - \ddot{a} a_o/ \dot{a} = 0.5 \sum \Omega_i(1+3w_i)$, when
$w_i$ is time independent.  

Thus, one way to determine the cosmological parameters 
$\Omega_i$ and the equations of state $w_i$, is to determine the coordinate 
distance to high-redshift sources, then use equations
given above, and solve for the cosmological parameters that 
yield the observed coordinate distances.  If the equation of
state is time-dependent, or if a rolling scalar field 
such as that proposed by Peebles \& Ratra (1988) 
is 
considered, then these equations must be modified accordingly
as discussed, for example, by Peebles \& Ratra (2002).    

The cosmological parameters determined using this method 
then go into the equation 
for $q_o$ to determine whether the universe is 
accelerating or decelerating today. The equation for
$q(z)$, which is very similar to that for $q_o$, then 
allows a determination of the redshift at which the 
acceleration is expected to go through zero, which marks the
redshift at which 
the universe transitions from acceleration to deceleration.  

A method of using the observed coordinate distances to go directly
to the acceleration parameter as a function of redshift will be
described below.  And, a method of using the data directly to
determine the function E(z) without making any assumptions about
the nature or redshift evolution of the Dark Energy will also be discussed.

The beauty of using coordinate distance measurements to determine 
the cosmological parameters $\Omega_i$ 
is that the true global cosmological parameters are determined; 
no corrections need to be made for the clustering properties of 
matter, so that contributions to $\Omega_i$ can not be missed or 
miscounted, and there are no biasing issues.

\section*{THE RADIO GALAXY AND SUPERNOVA METHODS}

Two methods currently being used to constrain global cosmological 
parameters through the determination of the coordinate distance 
to high-redshift sources are the Type Ia Supernova method 
(e.g. Perlmutter et al. 1999; Riess et al. 1998), and 
the Type FRIIb Radio Galaxy method (e.g. Daly \& Guerra 2002).    

For Type Ia Supernovae, there is one model parameter
$\alpha$, which goes into the determination of
$m^{eff}_B$,  the effective
apparent B band magnitude of the supernova at maximum brightness.
This is related to the Hubble-constant free luminosity distance
$D_L$ via the equation
\begin{equation}
m_B^{eff}(\alpha) = {\cal M_B} + 5 log (D_L [\Omega_i, w_i])~.
\end{equation}
If quintessence is being considered (e.g. Caldwell, Dave,
\& Steinhardt 1998), then $w_i$ would 
represent the equation of state, and if an evolving scalar
field is being considered, such as that proposed and studied
by Peebles and Ratra (1988), then $w_i$ would be evolving
with redshift (and would represent their parameter $\alpha$).   

${\cal M_B}$ is a constant obtained 
by fitting all of the data; it is related to the standard
absolute magnitude of the peak brightness of a supernova $M_B$:
${\cal M_B} = M_B +25 -5 ~log(H_o)$ (see Perlmutter et al. 1999).
For k = 0, and allowing for 
non-relativistic matter and quintessence,
there are (N-4) degrees of freedom; this is also the case for a 
universe with an evolving scalar field such as that discussed
by Peebles and Ratra (1988), or a universe with 
space curvature, a cosmological constant, and 
non-relativistic matter.

The Hubble-constant-free luminosity distance $D_L$ is related
to the dimensionless coordinate distance $y(z)$ via the 
relation $y(z)~(1+z) = D_L = H_o d_L$.

For FRIIb radio galaxies, there is one model parameter
$\beta$, which goes into the determination of the ratio
$R_* = <D>/D_*$ (Daly 1994).  This ratio also depends on the 
dimensionless luminosity distance $D_L$:
\begin{equation}
R*~(\beta,D_L[\Omega_i,w_i]) =  \kappa~.
\end{equation}

$\kappa$ is a constant obtained by fitting all of the 
radio galaxy data.    
This fit also has (N-4) degrees of freedom.
In these fits the dimensionless luminosity distance $D_L$,
and the dimensionless coordinate distance $y(z)$, is implicitly
determined for each source, though it only factors
out as a separate term when synchrotron cooling dominates
over inverse Compton cooling with 
CMB photons in the 
radio bridge of the source, in which case 
$ R* \equiv  <D>/D* = (observables)   (D_L)^{g(\beta)}$, 
where $g({\beta}) = 3/7 + 2\beta/3$   (Guerra \& Daly 1998).
This is not a valid approximation for all of the sources
in the sample.  
Here, this approximation has not been adopted, and an
iterative technique has been used to determine the dimensionless
coordinate distance to each source.  Thus, these coordinate distances
are valid for all of the sources in the sample.

The dimensionless luminosity distance $D_L$ and coordinate
distance $y(z)$ 
has been determined
using equation (3) for each source using the 
best fit value for
${\cal M_B}$, obtained with 
the 54 supernovae in the ``primary fit C'' of
Perlmutter et al. (1999) and the 1 high-redshift supernova
published by Reiss et al. (2001), with the magnitude
of this source corrected for gravitational lensing 
(Benitex et al. 2002). 
The
dimensionless coordinate distance $y(z)$ has been determined
for each radio galaxy 
using equation (4) and the 
best fit values for $\kappa$ and $\beta$ (along with 
their one sigma error bars)
for the
20 radio galaxies presented by Guerra, Daly,
\& Wan (2000).
These dimensionless coordinate distances are shown in Figures
1, 2, and 3.

\begin{figure}
\includegraphics[width=180mm]{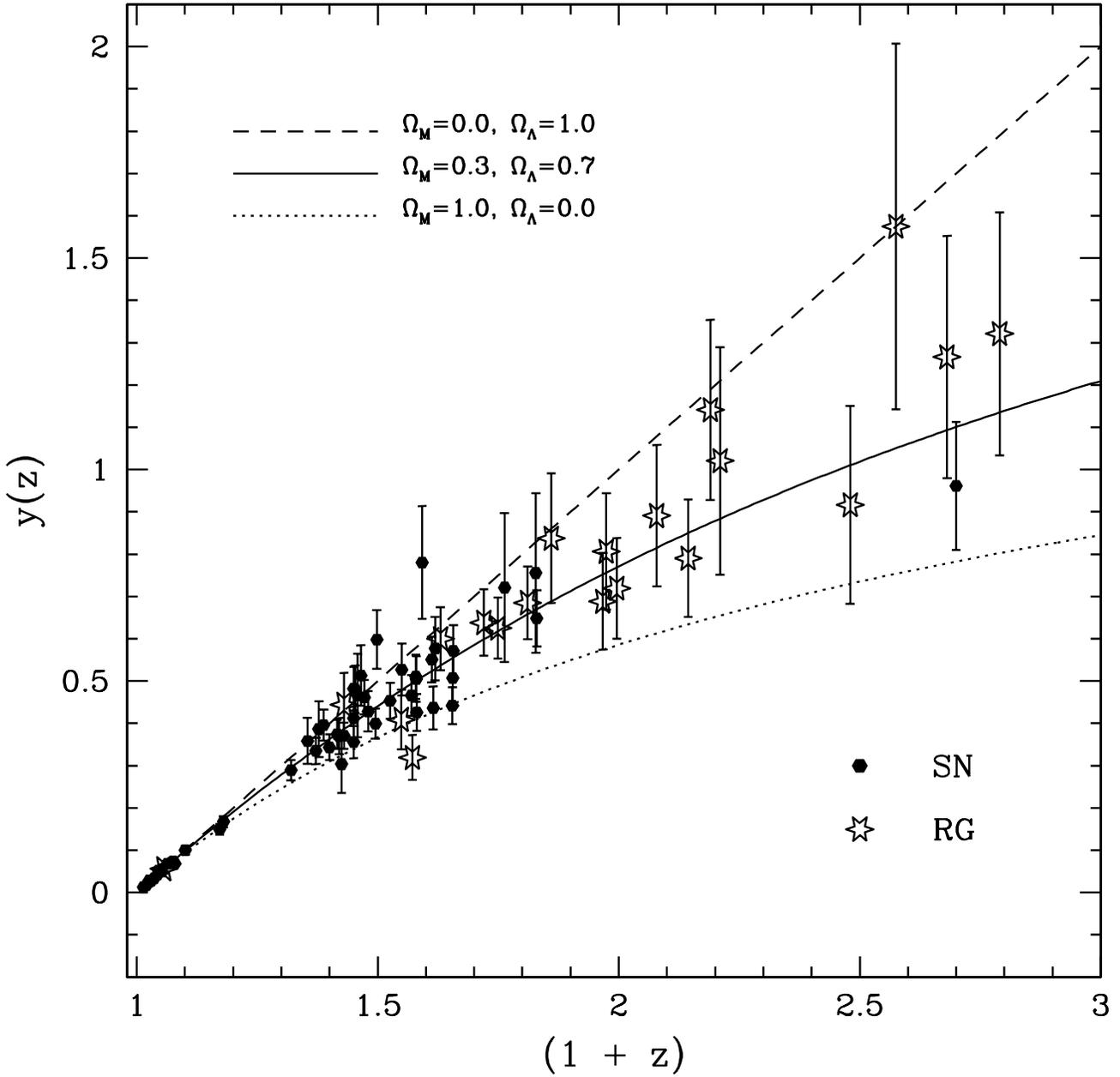}
\caption{Dimensionless coordinate distances $y(z)$ to 20 radio 
galaxies and 55 supernovae as a function 
of (1+z). Note that the radio galaxy and supernovae determinations
of $y(z)$ are {\sl completely independent}  
Radio galaxies are shown as open stars and
supernovae are shown as 
solid circles. }
\end{figure}

\begin{figure}
\includegraphics[width=180mm]{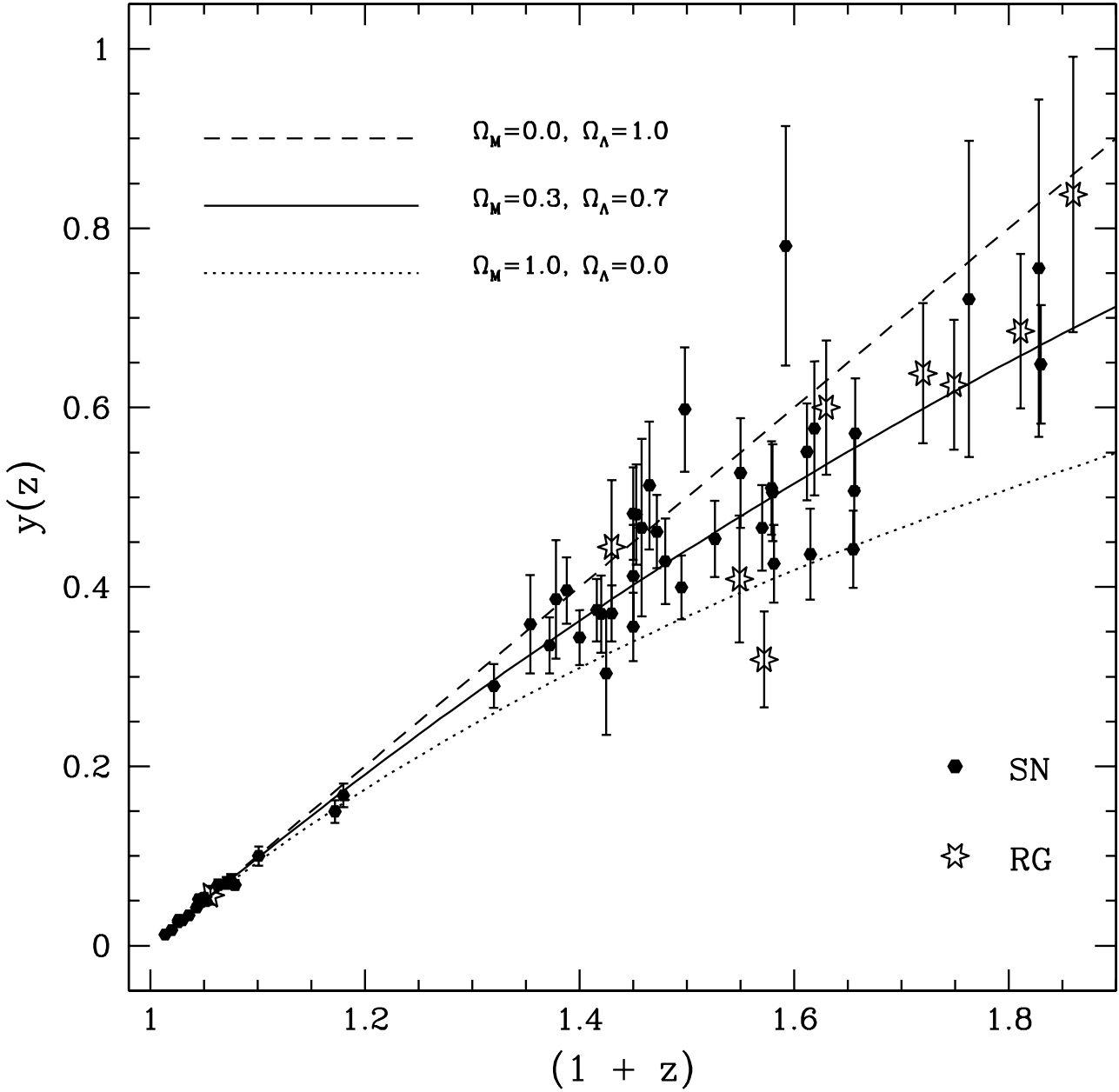}
\caption{Focus on the low redshift end of Figure 1.
Radio galaxies are shown as open stars and
supernovae are shown as 
solid circles. }
\end{figure}

\begin{figure}
\includegraphics[width=180mm]{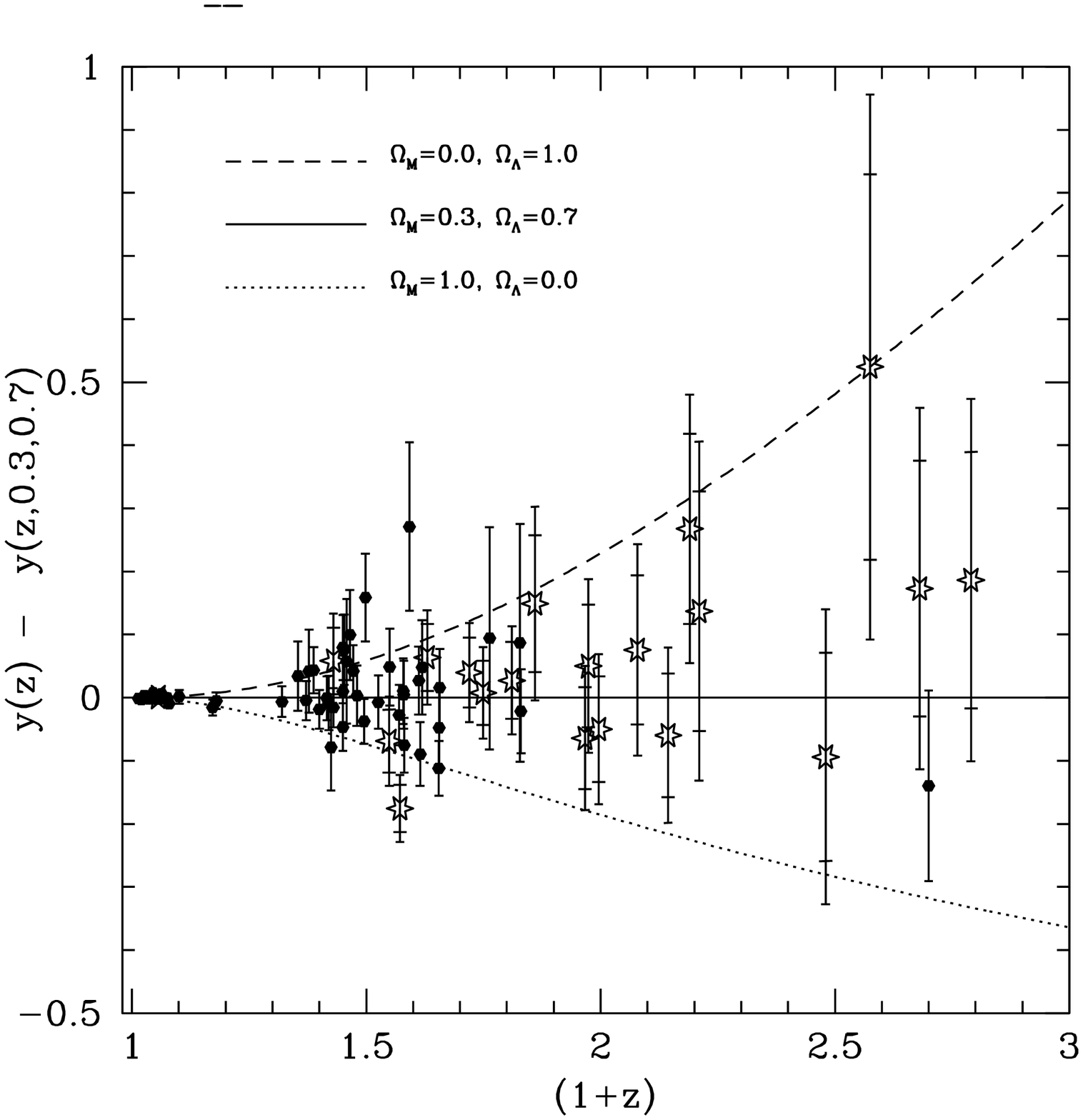}
\caption{The residuals between $y(z)$ and those expected in 
a universe with $\Omega_m = 0.3$
and $\Omega_{\Lambda} = 0.7$, where $y(z)$ is 
the dimensionless coordinate distance,
shown as a function of (1+z).  The error bars on the
radio galaxies could be reduced by a factor of 1.4 as indicated
on the figure if
the source 3C427.1 is excluded and the reduced chi-squared
is normalized to one, as discussed in detail by Podariu et al.
2003.  Radio galaxies are shown as open stars and
supernovae are shown as 
solid circles. } 
\end{figure}

The dimensionless coordinate distances obtained for the radio galaxies
are completely independent of those obtained for the supernovae.  
The conclusion from Figures 1, 2, and 3 
is that there is good agreement between results obtained using Type Ia SN and 
FRIIb RG.  This has also been demonstrated through detailed
analyses.  It is clearly the case when a cosmological constant
and non-relativistic matter are considered (Guerra, Daly, and Wan 2000), when 
quintessence is considered (Daly \& Guerra 2002), and
when the evolving scalar field model of Peebles \& 
Ratra (1988) is considered (Podariu, 
Daly, Mory, and Ratra 2002).

The results shown here, obtained using 20 FRIIb radio galaxies obtained 
from the published literature and the VLA archive, will be improved and 
extended with 10 new sources that will be observed 
by O'Dea, Guerra, Daly, \& Donahue at the VLA.    

Radio sources are observed out to very high redshift, and it would be easy to 
push this test to redshifts of three or four.  The only thing that would be 
required would be a week or two of observing time at the VLA, followed by 
data analysis.  This is not very demanding in terms 
of time, manpower, and funding.

\section{A MODEL-INDEPENDENT DETERMINATION OF q(z)}

The current method used to study the acceleration of the
universe is to take the measured $y(z)$ or $D_L(z)$, determine 
best-fitting global
cosmological parameters, and then use these global cosmological
parameters to determine the acceleration of the universe as
a function of redshift.  In this process, assumptions must
be made concerning the nature and redshift evolution of the 
mass-energy density of the ``dark energy." 

A direct, empirical determination of the 
acceleration of the universe as a function of redshift can be
determined using the data, without making any assumptions
about the nature or evolution of the ``dark energy.''
This can be done using the equation  
\begin{equation}
-q(z) \equiv \ddot{a} a/\dot{a}^2 = 1~ +~ (1+z) ~(dy/dz)^{-1} (d^2y/dz^2)
\end{equation}
valid for k=0; if $k \neq 0$, another term 
$[kr(1+z)/(1-kr^2)](dr/dz)$ must be 
added to the right hand side.   
Here, $y$ is the dimensionless coordinate distance
$y=H_o (a_or)$.   

Thus, equation (5) can be used to empirically determine the 
redshift at which the universe transitions from acceleration
to deceleration without requiring assumptions regarding
the nature and redshift evolution of the ''dark energy.''
The supernova and radio galaxy data allow a determination
of the dimensionless coordinate distance $y$ to each source,
at redshift $z$.  These data can then be used to determine
$dy/dz$, and $d^2y/dz^2$; these can then be substituted into
eq. (5) to determine q(z).  

Equation (5) follows from the RW line element and
the relation  $(1+z) = a_o/a(t)$. Our measurements
of the coordinate distance $(a_or)$ move along the
negative direction of $dr$, so eq. (2) with k=0 implies that 
$a_o~dr = - (1+z)~dt$, or   $(dz/dt) = -a_o^{-1}(1+z)~(dr/dz)^{-1}$.  
Differentiating $(1+z) = a_o/a(t)$ with respect to time   
implies that $\dot{a} = -a_o~(1+z)^{-2} ~(dz/dt)$.
Substituting in for $(dz/dt)$, we find
$\dot{a} = (1+z)^{-1} (dr/dz)^{-1}$.  Differentiating
again with respect to time, we find
$\ddot{a}=-(1+z)^{-2}~(dz/dt)~(dr/dz)^{-1}~[1+~(1+z)(dr/dz)^{-1}(d^2r/dz^2)]$,
which simplifies to eq. (5) using the expressions given here,
and the relation $y(z) = H_o (a_or)$.  
Note, these expressions include the well-known equation
(e.g. Peebles 1993; Weinberg 1972)
\begin{equation}
H(z) \equiv \dot{a}/a = \sqrt{1-kr^2}~~[d(a_or)/dz]^{-1}~,  
\end{equation}
or, for $k=0$ and with $H(z) = H_o E(z)$, 
\begin{equation}
E(z) = (dy/dz)^{-1}~.
\end{equation}

Since eqs. (5), (6), and (7) are derived without any assumptions regarding the
mass-energy components of the universe 
or their redshift evolution, they can 
be used to directly determine the 
function $E(z)$, which contains important information 
on the ``dark energy'' and its redshift evolution, and 
to determine the dimensionless 
acceleration parameter $q(z)$ directly from measurements
of $y(z)$. The use of the data to directly determine $E(z)$ and 
$q(z)$ is underway.

\section{DETAILS OF THE RADIO GALAXY METHOD}

The Radio Galaxy Method
relies on a comparison of the average size of an FRIIb source 
determined by the mean size $<D>$ of the full population at that redshift 
[$<D> \propto  (a_or)$], and the average source size, $D_*$, 
determined using a physical 
model that describes the evolution of the source. 
The two measures of the average source size should 
track each other, so $<D>/D_*$ should remain constant, independent
of redshift.  The ratio depends upon observed quantities, 
the coordinate distance and the 
model parameter $\beta$. When inverse 
Compton cooling with CMB photons is negligible, then 
$R_* \equiv  <D>/D_* = (Observables)(a_or)^{2\beta/3 + 3/7}$
(Guerra \& Daly 1998).
However, this is not a good approximation for all of the
sources in the radio galaxy sample.  Therefore, this approximation
has not been adopted here, and the full equation 
$R_* = k_o y^{(6 \beta -1)/7}~(k_1y^{-4/7}+k_2)^{(\beta/3-1)}$, where
$k_o,~k_1,$ and $k_2$ represent directly observed quantities,
has been used to solve for $y(z)$ to each source.

Details of the model are discussed by Daly \&
Guerra (2002), Guerra, Daly, \& Wan (2000), Guerra \& Daly (1998),
and Daly (1994).  The physics of FRIIb sources is discussed at
length by Daly (2002).  

In brief, the average size of a given source is
$D* = v_L t_*$, where $t_*$ is the total time that
the AGN produces large-scale jets.  Several different
lines of argument, reviewed by Daly \& Guerra (2002),
suggest that $t_* \propto L_j^{-\beta/3}$,
where $L_j$ is the beam power of the source.
If this relation is assumed, then $D_* \propto	
(B_L a_L)^{-2\beta/3}~  v_L^{1-\beta/3}\propto y^{-6\beta/7 + 8/7}~
(k_1 y^{-4/7}+k_2)^{1-\beta/3}$.  
For a typical value of 
$\beta$ of 1.7, $D_* \propto (B_L a_L)^{-1.1}~v_L^{0.4}$.     

Note that the length of the source, D, does not enter into our 
determination of D*.  Of course, D does enter into the determination of 
$<D>$. 

The determination of the most uncertain of these parameters, $v_L$,
can be studied using Chandra data, since Chandra data can be used
to determine or constrain the ambient gas density $n$, and 
the radio galaxy model assumes that $v_L^2 \propto B_L^2/n$.  
This is discussed below in the section ``Tests of the Radio Galaxy Model.'' 

The assumption that $t_* \propto L_j^{-\beta/3}$ is 
consistent with models of jet production via the electromagnetic 
extraction of the spin energy from a rotating black hole. Daly
\& Guerra (2002) show that this relation results if the 
magnetic field strength in the vicinity of the rotating
black hole is given by 
$B \propto (a/m)^{\beta/(3-\beta)}~ M^{(2\beta-3)/(2(3-\beta))}$,
or $B \propto (a/m)$ for $\beta = 1.5$, and
$B \propto (a/m)^2~M^{0.5}$ for $\beta=2$,  
where $a$ is the spin angular momentum per unit mass, $m$ 
is the gravitational radius of the black hole, and $M$ is the
mass of the black hole (see Blandford 1990).   

The two other key assumptions of the radio galaxy model,
that all FRIIb classical doubles at a given redshift 
have a similar maximum or average size so that the average 
size of a given source will be close to 
that of full population at that redshift, and that 
strong shock physics applies near the forward region of the radio source,
have been tested empirically, and are consistent with the data.

\section{TESTS OF THE RADIO GALAXY METHOD}

Three tests of the radio galaxy model and method are 
underway.  These are tests using Chandra X-ray data;
tests based on a detailed analysis of radio maps at multiple 
frequencies; and tests using a detailed comparison 
of radio maps with results from numerical simulations.
				
Chandra X-ray data can be used to study the ambient gas density, 
which can be compared with that predicted.  
Donahue, Daly, and Horner (2002) have completed Chandra studies of 
3C 280 (a radio galaxy) and 3C 254 (a radio loud quasar).  The 
predicted ambient gas densities are consistent with the 
three sigma upper bounds placed by the Chandra data.  Other radio predicted 
ambient gas densities will be compared with Chandra observations and bounds.

Ten new RG with redshifts between 0.4 and 1.3 
will be observed with the VLA 
in collaboration with Chris O'Dea, Eddie Guerra, and Megan Donahue.  
These high-resolution observations will allow several tests of the RG model, 
and will be compared with detailed numerical simulations tailored to match 
the properties of these sources, which is being done in collaboration with 
Joel Calvalho and Chris O'Dea.  These studies will be based on 
the numerical work of Calvalho \& O'Dea (2002a,b), and  will 
help to identify the physical processes that must be accounted for 
the in modeling FRIIb sources.

In addition, complete radio galaxy samples of FRIIb 
sources that go to redshifts of three or four will be investigated.  
These could serve as the parent populations 
in the application of the RG method to redshifts greater than two.
This could take us to very high redshift very quickly since
radio observations are relatively quick, easy, and
do not require major new instruments.  

\section{CONCLUSIONS}  

In a spatially flat universe with 
non-relativistic matter and quintessence, 
radio galaxies alone indicate with 
84\% confidence that the universe is accelerating in its expansion at the 
present epoch (Daly \& Guerra 2002).  
Results obtained using the 
Radio Galaxy Method out to redshifts of two are consistent 
with those obtained using the Supernova Method out to 
redshifts of about one. 

A model-independent way to 
use the supernova and radio galaxy data to determine
the dimensionless expansion rate $E(z)$ and 
acceleration parameter $q(z)$ is presented and 
discussed.  A direct determination of $q(z)$ that is
independent of assumptions regarding the nature and
evolution of the ''dark energy'' would allow the
transition redshift from acceleration to deceleration
to be determined, and would help to identify any  
systematic errors that might plague either the 
radio galaxy or supernova methods.   A direct determination
of $E(z)$ would help to quantify the properties and 
redshift evolution of the ``dark energy'' and to identify
potential systematic errors in the methods used to 
determine $y(z)$.

\section*{ACKNOWLEDGEMENTS}

It is a pleasure to thank Joel Carvalho, 
George Djorgovski, Megan Donahue, Jean Eilek, 
Eddie Guerra, Paddy Leahy, 
Alan Marscher, Matt Mory, Chris O'Dea, 
Bharat Ratra, and Adam Reiss for 
helpful comments and discussions.  This work was
supported in part by US National Science Foundation
grants AST 00-96077 and AST 02-06002, by 
a Chandra X-ray Center
data analysis grant G01-2129B, and by Penn State
University.  
The Chandra X-ray Observatory Science
Center (CXC) is operated for NASA by the Smithsonian Astrophysical
Observatory.

\end{document}